\documentstyle[psfig]{mn0}

\def\simg{\mathrel{\rlap{\raise 0.511ex \hbox{$>$}}{\lower 0.511ex \hbox{$\sim$}}}}
\def\siml{\mathrel{\rlap{\raise 0.511ex \hbox{$<$}}{\lower 0.511ex \hbox{$\sim$}}}}
\def\beq{\begin{equation}} \def\eeq{\end{equation}}

\begin{document}

\title[Early optical afterglows]{Taxonomy of GRB optical light-curves:
       identification of a salient class of early afterglows}

\author[Panaitescu \& Vestrand]{A. Panaitescu and W.T. Vestrand \\
       Space Science and Applications, MS D466, Los Alamos National Laboratory,
       Los Alamos, NM 87545, USA}

\maketitle

\begin{abstract}
 The temporal behaviour of the early optical emission from Gamma-Ray Burst afterglows 
can be divided in four classes: fast-rising with an early peak, slow-rising with a 
late peak, flat plateaus, and rapid decays since first measurement. The fast-rising 
optical afterglows display correlations among peak flux, peak epoch, and post-peak 
power-law decay index that can be explained with a structured outflow seen off-axis, 
but the shock origin (reverse or forward) of the optical emission cannot be determined. 
The afterglows with plateaus and slow-rises may be accommodated by the same model, 
if observer location offsets are larger than for the fast-rising afterglows, or 
could be due to a long-lived injection of energy and/or ejecta in the blast-wave. 
If better calibrated with more afterglows, the peak flux -- peak epoch relation 
exhibited by the fast and slow-rising optical light-curves could provide a way to use 
this type of afterglows as standard candles.
\end{abstract}


\section{Introduction}

 The ability of the Swift satellite to localize precisely Gamma-Ray Bursts 
(GRBs) in real-time has allowed multi-wavelength monitoring of bursts starting 
at tens of seconds after trigger, even before the end of the prompt gamma-ray 
emission phase. The new observations have shown that the optical emission from 
GRBs can be divided into two components: a counterpart emission that tracks the 
prompt gamma-rays and an afterglow emission that starts during the prompt phase 
or shortly after it and which dims progressively for hours or days.
The former component was observed for the first time in GRB 041219A (Vestrand et 
al 2005), the latter was displayed by GRB afterglow 990123 during the burst 
(Akerlof et al 1999) and by GRB afterglow 030418 after the burst (Rykoff et al 2004).
Both components are seen simultaneously in the prompt optical emission of GRB 
050820A (Vestrand et al 2006).

 As of the end of 2007, there are about 30 GRBs with known redshift and whose 
optical afterglow emission was monitored starting within a few minutes after the 
trigger. This set of well-sampled optical light-curves is large enough 
for a collective study of their properties, as was done by Akerlof \& Swan (2007)
and Kann et al (2007) for the early (0.1--1 ks) afterglow emission and by Liang 
\& Zhang (2006), Nardini et al (2006), and Zeh, Klose \& Kann (2006) at 1 d.

 In this work, we examine the temporal properties of the early optical afterglow 
emission, identify a peculiar class of afterglows with initially rising optical
light-curves, whose properties render them standard candles, and attempt to explain 
those properties in the standard theoretical framework of a relativistic blast-wave 
interacting with the circumburst medium (e.g. Paczy\'nski \& Rhoads 1993, 
M\'esz\'aros \& Rees 1997).

\begin{table}
\caption{ Properties of the early optical afterglows used in this work 
   (their $z=2$ light-curves are shown in Figure \ref{z2}). }
\begin{tabular}{lcccccccccccc}
  \hline \hline
    GRB &   z   &  $F_p$  &  $t_p$  & $\alpha_o$ & Refs \\
        &       &  (mJy)  &   (s)   &            &      \\
        &  (1)  &  (2)    &   (3)   &   (4)      &  (5) \\
  \hline \hline
  \multicolumn{2}{l}{\sc fast-risers} \\
 990123 &  1.60 & 1240(R) &   50    &    1.80    & A99,G99 \\
 050730 &  3.97 &  1.6(R) &   550   &    0.63    & P06,P07 \\
 050820A&  2.61 &  4.6(R) &   420   &    0.91    & C06,V06 \\
 060418 &  1.49 &  52(H)  &   120   &    1.13    & Mo07 \\
 060607A&  3.08 &  17(H)  &   160   &    1.20    & Mo07 \\
 061007 &  1.26 &  500(R) &    57   &    1.70    & Mu07,Y07b \\
  \hline
  \multicolumn{2}{l}{\sc slow-risers} \\
 060614 &  0.12 & 0.096(R)&  21000  &    1.54   & DV06,Ma07 \\
 060904B&  0.70 & 0.92(R) &   520   &    0.88   & K08 \\
 070110 &  2.35 & 0.065(V)&   5000  &    0.65   & T07 \\
 070411 &  2.95 & 0.18(R) &$\sim700$&    0.94   & 6269-88-95\\
 071025 &$\sim4$&  1.4(R) &   570   &    2.0    & 7008-11-18 \\
  \hline
  \multicolumn{4}{l}{\sl fast or slow-riser (uncertain)} \\
 050904 &  6.29 & 4--10(J)&$\sim430$&    1.15   & T05,Bo06,H06 \\
  \hline \hline
  \multicolumn{2}{l}{\sc decays} \\
 021211 &  1.01 &  3.9(R) &   110   &    1.58   & F03a,L03 \\
 050319 &  3.24 &  1.3(R) &   170   &    0.88   & W05,Q06 \\
 050401 &  2.90 & 0.69(R) &    35   &    0.80   & R05,dP06,W06 \\
 050416A&  0.65 & 0.078(R)&   115   &    0.54   & H07,S07 \\
 050525 &  0.61 &   23(R) &    66   &    1.09   & K05,Bl06 \\
 050908 &  3.34 & 0.082(R)&   380   &    0.83   & 3944-45-47-50 \\ 
 050922C&  2.20 &  6.3(R) &   160   &    0.70   & 4040-41-95 \\
 051109A&  2.35 &  3.4(R) &    40   &    0.67   & 4239,Y07a \\
 051111 &  1.55 &  24(R)  &    32   &    0.80   & Bu06,Gu07,Y07a \\
 060927 &  5.47 &  4.2(I) &    20   &    1.01   & RV07 \\
 061121 &  1.31 &  0.97(R)&    46   &    0.61   & 5847,Y07b \\
 061126 &  1.16 &  59(R)  &    23   &    1.34   & P08,Y07b \\
  \hline \hline
  \multicolumn{2}{l}{\sc plateaus} \\
 021004 &  2.30 &  2.8(R) &   350   &    0.28   & F03b,H03 \\
 050801 &  1.56 &  4.1(R) &    25   &    0.13   & 3726-33,dP07 \\
 060124 &  2.30 &  1.0(R) &   180   & $\sim 0$  & R06,Mi07 \\
 060210 &  3.91 &  0.19(R)&    95   &    0.13   & C07 \\
 060714 &  2.71 &  0.18(R)&   300   &    0.17   & 5434,K07 \\
 060729 &  0.54 & 0.34(W1)&    78   &    0.15   & Gr07,Y07b \\
 \hline \hline \\
\end{tabular}
\begin{minipage}{85mm}
 {\bf (1)}: burst redshift, $z=4$ was assumed for GRB 071025, as it was a $V$-band dropout
       (GCN 7011), which implies a photometric redshift $z\in(3.5,4.3)$;
 {\bf (2)}: peak flux for fast- and slow-rising afterglows or flux at the first measurement for
      the other types; corrected for Galactic dust extinction and in the optical band indicated
      in parenthesis; $1\sigma$ uncertainty of peak flux is usually 5--15 percent;
 {\bf (3)}: epoch of the optical light-curve peak for fast and slow risers or epoch of first
       measurement for the rest; $1\sigma$ uncertainty of peak time is 10--40 percent;
 {\bf (4)}: index of the optical flux power-law decay during the early afterglow
       ($F_o \propto t^{-\alpha_o}$); for rising afterglows, this is the decay after the
       peak; uncertainties are around 0.05;
 {\bf (5)}: References for optical data used in this article -- 
     GCN Circulars: 3726 (Rykoff et al), 3733, (Blustin et al), 3944 (Cenko et al), 3945 (Li),
     3947 (Kirschbrown et al), 3950 (Durig et al), 4040 (Fynbo et al), 4041 (Hunsberger et al),
      4095 (Li et al), 4239 (Wo\'zniak et al), 5434 (Asfandyarov et al),
     5847 (Halpern et al), 6269 (Rykoff et al), 6288 (Mikuz et al), 6295 (Kann et al),
     7008 (Wren et al), 7011 (Milne \& Williams), 7018 (Minezaki et al). 
\end{minipage}
\end{table}

\section{Rising optical light-curves}

 The GRB afterglows with early optical coverage used in this work are listed in Table 1.
 To facilitate the comparison of the light-curve properties, we have $k$-corrected 
the observed optical flux $F(\nu,t)$ of those GRBs to a fiducial redshift of $z_0=2$ and 
same observing frequency $\nu_0 = 4.8 \times 10^{14}$ Hz (corresponding to 2 eV, i.e. 
the $R$-band)
\begin{equation}
 F \left[ \nu_o,\frac{z_0+1}{z+1} t \right] = \left[ \frac{d_L(z)}{d_L(z_0)} \right]^2
          \hspace*{-2mm} \left( \frac{z_0+1}{z+1} \right)^{ \hspace*{-1mm} 1-\beta_o}
          \hspace*{-1mm} \left( \frac{\nu_0}{\nu} \right)^{ \hspace*{-1mm} -\beta_o} 
                     \hspace*{-1mm} F(\nu,t)
\end{equation}
where $z$ is the burst redshift, $d_L$ the luminosity distance, assuming an optical 
spectrum $F_\nu \propto \nu^{-\beta_o}$ with $\beta_o = 0.75$, which is typical for 
the optical emission at $\sim 1$ day of pre-Swift afterglows. We note that an error 
of 0.25 in the optical spectral slope yields an error of less than 20 percent in the 
$k$-corrected optical flux.

 Optical fluxes were corrected for Galactic extinction. Evidence for a substantial 
host extinction exists for GRB afterglows 050401 (Watson et al 2006), 050525 (Blustin 
et al 2006), 060927 (Ruiz-Velasco et al 2007), 061007 (Mundell et al 2007, Schady et 
al 2007), and 061121 (Page et al 2007), whose optical spectrum is $F_\nu \propto 
\nu^{-\beta}$ with $\beta \in (1.5,2.4)$, redward of intergalactic Ly$\alpha$ absorption. 
By assuming a simple reddening curve for the dust in the host galaxy, $A_\nu \propto \nu$, 
we have estimated the host extinction $A_V$ in the host-frame $V$-band from the observed 
spectral slope $\beta$ and the assumed intrinsic slope (0.75): $A_V \simeq
(\beta-0.75)/(z+1)$, and have correct the optical fluxes of GRB afterglows 050401, 
050525, 060927, 061007, and 061121 for host-extinction of $A_V \simeq$ 0.3, 0.6, 0.2, 
0.5, 0.3 mag, respectively.

\begin{figure*}
\psfig{figure=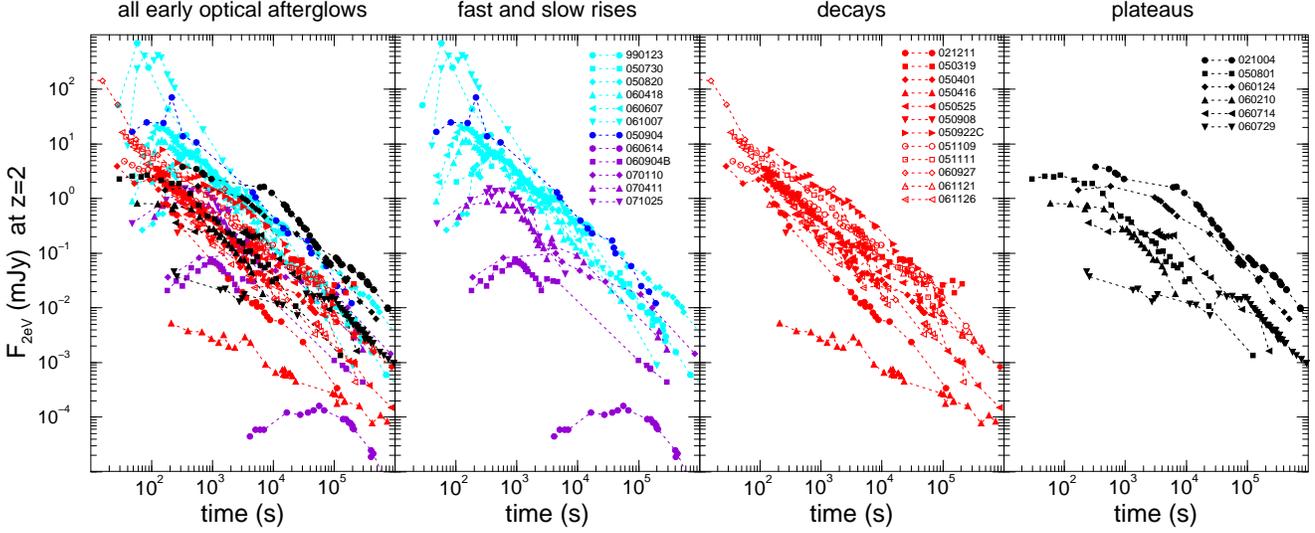,width=17.5cm}
\caption{
    Optical light-curves of 28 GRB afterglows with known redshift, most of which
    were followed starting 1--few minutes after trigger.
    Optical fluxes have been corrected for Galactic and host dust extinction (the latter 
    being estimated from the observed optical spectral slope and assuming an intrinsic 
    slope of 0.75) and calculated for a common redshift $z=2$.
    Color coding: light-blue for 6 afterglows with a fast rise, purple for 5 slow risers,
    dark-blue for GRB 050904 of uncertain type (fast or slow-rise), red for 12 afterglows 
    with a decay since first observation (i.e. their peaks occurred earlier than first 
    measurement and have been missed), black for 6 afterglows with optical plateaus.
    Note that the luminosity of the afterglows with fast rises has a very narrow
    distribution at 0.5--5 ks, although they peak at different times. The other types
    of optical afterglows (plateaus and decays) have much wider luminosity distributions.
}
\label{z2}
\end{figure*}

 The resulting optical light-curves are shown in Figure \ref{z2}. Even after 
redshift and host extinction corrections, the optical fluxes at any fixed 
time span a range of at least 2--3 orders of magnitude, displaying a well-defined
boundary on the bright side which, obviously, is not an observational selection effect.
Based on the optical light-curve behaviour at $30-10^4$ s after trigger, the afterglows
shown in Figure \ref{z2} can be separated in four groups: \\
(i) 6 fast-rising ($F_o \propto t^{2.5\pm 0.5}$), peaking at about 100 s, \\
(ii) 5 slow-rising ($F_o \propto t^{0.6\pm 0.2}$), peaking after 100 s, \\
(ii) 12 with fast decays ($F_o \propto t^{-1.0\pm 0.3}$) since first measurement
      (at about 100 s), \\
(iv) 6 with plateaus ($F_o \propto t^{-0.2\pm 0.1}$), lasting for 1--2 decades in time.

 Evidently, the last two types of afterglow had a fast or slow rise to their peaks, which
occurred before the first observation. The afterglows with plateaus represent a separate
class based on their slow post-peak decay, however the afterglows with decays have post-peak
power-law decays ($F_o \propto t^{-\alpha_o}$) of exponents comparable with those of the
slow and fast-rising afterglows (Table 1), hence their classification stems (so far) only
from that their peak epochs were earlier than for the rising afterglows.

 The fast-rising afterglows define a family of curves (shown with light-blue lines in 
Figure \ref{z2}) that roughly delineate the high-brightness boundary mentioned above.
 In addition to peaking sufficiently late (around 100 s after trigger), thereby
allowing robotic telescopes to catch their rise, the fast-rising afterglows display
other peculiar features which suggest that they represent a distinct class. As shown
in Figure \ref{z2}, their luminosity distribution is substantially narrower than for 
the other types of afterglows. 

\begin{figure}
\psfig{figure=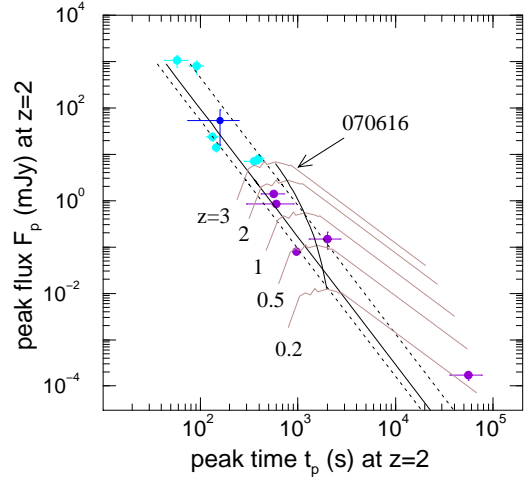,height=6.5cm,width=7cm}
\caption{ 
   Correlation of $z=2$ optical light-curve peak flux (at 2 eV) and peak time for 
   6 afterglows with fast rises (light-blue dots) and 5 afterglows with slow rises 
   (purple dots).
   Solid straight line shows the best fit given in equation (\ref{peak}) and the dotted 
   lines its $1\sigma$ uncertainty (i.e. $\sim 2/3$ of points are within the dotted lines).
   The probability of obtaining that correlation in the null hypothesis is $10^{-5.5}$ 
   ($4.7\sigma$ significance level).
   GRB afterglow 050904 is shown by the dark-blue dot.
   Faded solid lines show the $z=2$ optical light-curve of GRB afterglow 070616,
   whose redshift is not known, assuming various burst redshifts (black solid curve 
   connects the peaks). For $z\in (0.2,1)$, the peak of the afterglow 070616 optical 
   light-curve falls within the $1\sigma$ uncertainty of the $F_p-t_p$ relation. 
   (Optical data for 070616 are from Starling et al 2007 and GCN 6547 (Fatkhullin et al) 
   and have been corrected for Galactic extinction of $E(B-V)=0.4$). 
  }
\label{Fptp}
\end{figure}

 Figure \ref{Fptp} illustrates another intriguing feature of the fast-rising optical
afterglows: an anti-correlation of the peak flux ($F_p$) and peak epoch ($t_p$) with
a linear correlation coefficient $r(\log F_p, \log t_p) = -0.88 \pm 0.04$, corresponding
to less than 3 percent probability to obtain by chance a correlation stronger than that.
 The 5 afterglows with slow-rises are consistent with that correlation; if added, 
we obtain for 11 afterglows with rising (slow or fast) optical light-curves that
$r(\log F_p, \log t_p) = -0.97 \pm 0.01$ (probability of a stronger chance correlation 
is less than $10^{-5.5}$)\footnotemark and a best fit
\begin{equation}
 \log (F_p/{\rm mJy}) = (7.5 \pm 0.5) - (2.7 \pm 0.2) \log (t_p/{\rm s})  
\label{peak}
\end{equation}
where the $1\sigma$ uncertainties of the two coefficients were calculated for the joint 
variation of them.
\footnotetext{Without GRB 060614, whose latest and dimmest peak lies farther from the 
  other 10 afterglows (see Figure \ref{Fptp}), $r(\log F_p, \log t_p) = -0.95 \pm 0.02$
  and the chance correlation probability is less than $10^{-4}$}

 If confirmed and better calibrated with a larger set of rising optical afterglows, 
this correlation may allow the estimation of afterglow redshifts based on properties 
of the optical light-curve. As shown in Figure \ref{Fptp}, the current uncertainty of
the peak flux -- peak time relation (dotted lines) is a bit too large to set a strong 
constraint on the burst redshift: the peak of the optical light-curve of GRB afterglow 
070616 falls within the $1\sigma$ uncertainty of the $F_p-t_p$ relation for $z\in (0.2,1)$.

 Another feature of the fast-rising afterglows which may prove to be a "standard candle"
is their luminosity at 0.5--5 ks. At $z=2$, the $R$-band optical flux of 5 of the 6 
afterglows with such rises is $F_{2eV} = (1.5-3) (t/1\,{\rm ks})^{-1.2}$ mJy, 
which corresponds to a source-frame $R$-band luminosity
\beq
 (\nu L_\nu)_{2eV} = (1.5-3) \times 10^{46} \left( \frac{t}{1\,{\rm ks}} \right)^{-1.2}
    \; {\rm erg\; s^{-1}} \;.
\eeq
Given the large spread in peak flux for the fast-rising afterglows, the relatively narrow
distribution of their optical luminosity at 1 ks arises from the post-peak power-law
decay index $\alpha_o$ being correlated with the peak flux: $r(\log F_p, \alpha_o) = 0.96 
\pm 0.02$ (less than 0.5 percent chance correlation). 

 It remains to be tested with a larger sample of afterglows if the rising afterglows 
represent a standard candle through either their $F_p-t_p$ anticorrelation or the
optical luminosity at $\sim 1$ ks of the afterglows peaking before 1 ks. 
For the remainder of this article, we attempt to provide an explanation for the above 
properties of the rising optical light-curves. To that end, we search for a model that 
accommodates the following features: \\
(i) a fast $F_o \propto t^{2-3}$ optical rise, \\
(ii) a peak flux of 1--1000 mJy at peak time 0.1--1 ks, \\
(iii) a later peak epoch for a smaller peak flux, \\
(iv) a faster post-peak decay for a brighter peak.

\section{Afterglow blast-wave models}

 The power-law decay displayed by GRB afterglow light-curves suggests that they
originate from a relativistic blast-wave decelerated by its interaction with the
ambient medium. The continuous transfer of energy to the swept-up medium decreases
the blast-wave's Lorentz factor as a power-law with radius which, together with 
the power-law energy spectrum of particles accelerated at shocks (either the
{\sl reverse-shock} propagating in the incoming ejecta or the {\sl forward-shock}
energizing the ambient medium), yield power-law decaying light-curves ($F_o \propto
t^{-\alpha_o}$) without any additional assumptions for the forward-shock emission,
but requiring an extra feature (a power-law ejecta mass distribution with Lorentz
factor) for the reverse-shock emission.

 We identify features of the forward- and reverse-shock models that may account 
for the properties of the fast-rising afterglows and the diversity of early
optical light-curves by calculating numerically the afterglow emission.  
The numerical model has the following components: \\
$(1)$ calculation of the dynamics of the two shocks, with allowance for the angular
    distribution of the ejecta kinetic energy \& initial Lorentz factor and for
    energy injection. After deceleration, the dynamics of the forward-shock is the
    Blandford-McKee solution. The Lorentz factor of the reverse-shock, as measured
    in the frame of the incoming ejecta, is determined from the Lorentz factor of
    those ejecta and that of the shocked gas, \\
$(2)$ setting the electron power-law distribution with energy, taking into account
    radiative losses, which yield a cooling break frequency. The total electron energy
    is quantified by the fraction $\varepsilon_e$ of the post-shock energy imparted
    to them. Similarly, the magnetic field energy is quantified by the fractional
    energy $\varepsilon_B$ stored in it, \\
$(3)$ calculation of the peak synchrotron flux from the number of radiating electrons
    behind each shock and the magnetic field strength, and of the spectral breaks:
    $(i)$ self-absorption frequency, $(ii)$ characteristic synchrotron frequency
    for typical electron energy, $(iii)$ cooling frequency (including also
    inverse-Compton cooling of electrons), \\
$(4)$ integration of the synchrotron emission over the evolution of each shock,
    taking into account geometrical-curvature and relativistic effects.
The theoretical formalism describing the above calculations can be found in e.g.
M\'esz\'aros \& Rees (1997), Sari, Piran, \& Narayan (1998), Wijers \& Galama (1999),
Panaitescu \& Kumar (2000, 2001).

\subsection{Forward-shock emission}

\subsubsection{Isotropic outflow -- pre-deceleration phase}

\begin{figure*}
\centerline{\psfig{figure=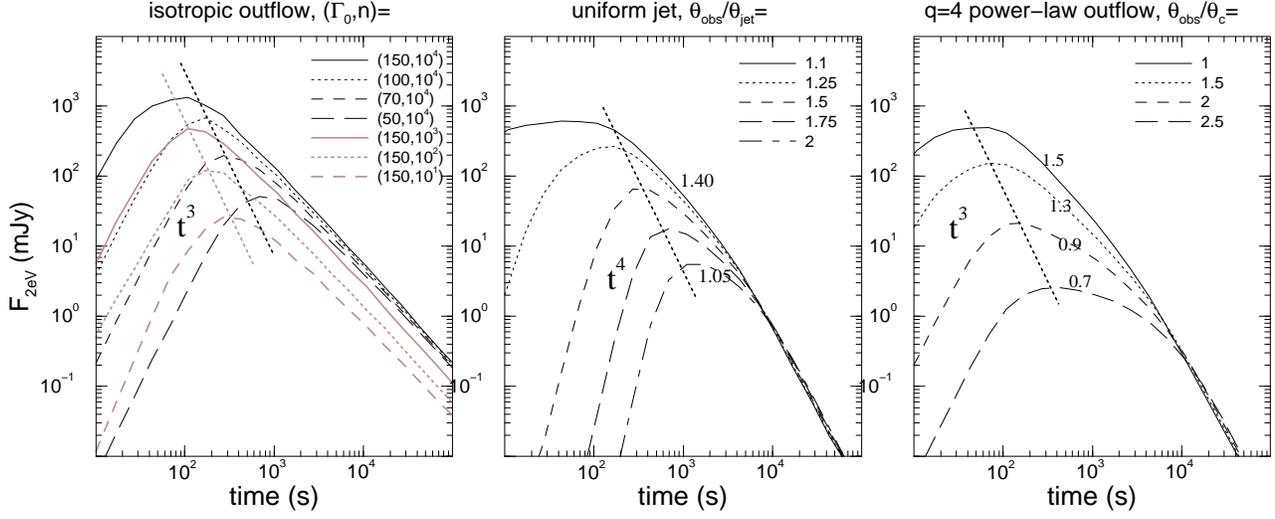,width=17cm}} 
\caption{ 
  Optical light-curves from the synchrotron emission of the {\bf forward-shock}
  driven into a {\bf homogeneous} circumburst medium by a relativistic outflow.
  To obtain a fast post-peak light-curve decay, the characteristic synchrotron
  frequency at which the shock-accelerated electrons radiate is placed just below
  optical at 100 s by choosing a magnetic field fractional energy $\epsilon_B = 0.01$
  and electron fractional energy $\epsilon_e = 0.003$.
  The outflow's kinetic energy per solid angle is ${\cal E}_0 = 10^{54}\,{\rm erg\, sr^{-1}}$.
  The slanted line indicates the slope of the $\log F_p - \log t_p$ fit to observations
  shown in Figure \ref{Fptp}.
 {\bf Left panel}: Pre-deceleration, rising light-curves for an isotropic outflow
  (kinetic energy per solid angle does not change with direction) and a semi-relativistic
  reverse-shock. The forward-shock light-curve peak marks the onset of deceleration,
  whose epoch depends on the ejecta's initial Lorentz factor $\Gamma_0$ and medium
  density $n$ (given in legend). The flux rise prior to the onset of deceleration
  is due to the fast increase in the number of ambient medium electrons swept-up and
  energized by the forward-shock.
  The post-peak power-law light-curves $F_{2eV} \propto t^{-\alpha_o}$ have the same
  index $\alpha_o$ for any peak time $t_p$ or peak flux $F_p$, unlike the correlation
  exhibited by the fast-rising optical afterglows.
 {\bf Middle panel}: Rising optical light-curves can also be obtained with a uniform
  jet (outflow of constant kinetic energy per solid angle ${\cal E}$ within half-opening
  $\theta_{jet}$) for various angles $\theta_{obs} > \theta_{jet}$ (given in legend)
  between the jet axis and the observer's line of sight ($\theta_{obs}=0$ for a jet
  moving exactly toward the observer). Here $\theta_{jet} = 1$ deg, thus jet energy
  is $E_{jet} = 10^{51}$ erg. To match the observed peak optical flux, a high ambient
  density ($n=10^4\, {\rm cm^{-3}}$) was used.
  The initial Lorentz factor was chosen sufficiently large to ensure that deceleration
  starts well before the earliest time shown.
  In this model, the fast rise of the light-curve is caused by that, as the jet
  decelerates, its emission is progressively less beamed relativistically off the
  direction toward the observer.
 {\bf Right panel}: A stronger $t_p-\alpha_o$ dependence is obtained with an outflow
  endowed with angular structure and various observer locations.
  Here, we used a power-law outflow, whose kinetic energy per solid angle changes
  with direction measured from the symmetry axis as ${\cal E} (\theta) = {\cal E}_0
  (\theta/\theta_c)^{-4}$. The core, of angular size $\theta_c = 1$ deg,
  is considered uniform. Legend gives the observer location in units of $\theta_c$.
  The initial Lorentz factor distribution is $\Gamma_0 \propto {\cal E}^{1/2}$
  (which leads to a deceleration radius $r_d \propto ({\cal E}/\Gamma_0^2)^{1/3}$
  that is angle-independent) with $\Gamma_0 = 200$ on the symmetry axis.
  The circumburst medium density is $n=10^4\, {\rm cm^{-3}}$.
  In this model, the light-curve's fast rise is the pre-deceleration emission
  from the region moving toward the observer, while the slower post-peak decay
  arises from the more energetic outflow core becoming visible to the observer.
  The post-peak decay indices $\alpha_o$ are indicated; their range is similar to
  that measured for the fast-rising optical light-curves of Figure \ref{z2}.
  {\sl Middle and right panels}: Compatibility of model light-curves peak fluxes
  and epochs with the observed relation (slanted lines) is a test of outflow
  properties universality (the only the changing parameter is observer location
  relative to the outflow's symmetry axis).
 }
\label{s0jet}
\end{figure*}

 An isotropic outflow or a jet seen face-on can yield rising (synchrotron emission)
light-curves after the forward-shock deceleration starts, however a fast rise
($F_\nu \propto t^{1-2}$) is obtained only at observing frequencies below the
synchrotron self-absorption frequency $\nu_a$, the next fast rise at $\nu > \nu_a$
being $F_\nu \propto t^{1/2}$, which is too slow to explain the rising optical afterglows.
The simplest test of optical being below $\nu_a$ is to measure the slope of the spectral
energy distribution ($F_\nu \propto \nu^{-\beta}$) of early optical afterglows.
This test requires multiband follow-up of the optical afterglow during the light-curve
rise, i.e. during the first few minutes after trigger. Few such observations are available:
for GRB afterglow 060418, we find that $\beta = 0.9 \pm 0.9$ before the light-curve
peak epoch while for GRB afterglow 060607A $\beta = 0.8 \pm 0.2$ at the peak epoch.
These spectral slopes are inconsistent with the $F_\nu \propto \nu^2$ expected below
$\nu_a$, hence the fast-rising optical afterglows do not arise from an isotropic,
decelerating outflow. Furthermore, the measured early optical spectral slopes indicate
that the optical domain is above the peak of the spectrum as early as a few hundred
seconds after trigger.

 Instead, a fast-rising forward-shock emission light-curve can be obtained before the
onset of that shock's deceleration, defined by the reverse-shock crossing the shell of
relativistic ejecta.

 For an comoving-frame ejecta density smaller than $4\Gamma_0^2 n$, where $\Gamma_0$
is the ejecta initial Lorentz factor and $n$ the ambient medium proton density, the
reverse-shock is relativistic and the shell-crossing time is set by the thickness
of the ejecta. In this case, if the ejecta are uniform (i.e. their density does not
vary with geometrical depths) the forward-shock Lorentz factor decreases slowly before
the reverse-shock crosses the ejecta shell, leading to a slowly rising forward-shock
light-curve: $F_o \propto t^1$ for a homogeneous medium and $F_o \propto t^{1/9}$ for
a wind medium.

 If the comoving-frame ejecta density is larger than $4\Gamma_0^2 n$, the reverse-shock
is semi-relativistic and, when the reverse-shock crosses the ejecta shell, the
forward-shock has swept-up an ambient medium mass equal to a fraction
$1/\Gamma_0$ of the ejecta mass, hence the deceleration timescale is set by the
ejecta Lorentz factor. If the ejecta are uniform then, prior to deceleration,
the forward-shock moves at constant Lorentz factor, owing to the continuous energy
input from the incoming yet-unshocked ejecta. For a homogeneous medium, the
pre-deceleration forward-shock light-curve rises as $F_o \propto t^2$ or $F_o \propto
t^3$ (depending on the location of the cooling frequency), exactly as observed for
fast-rising afterglows. For a wind medium, the fastest possible rise at a frequency
above the peak of the synchrotron spectrum (as implied by the optical spectral slope
of GRB afterglows 060418 and 060607A) is ($F_o \propto t^{1/2}$), which is too slow
compared to the fast-rising optical afterglows.

 Thus, if the ejecta shell is uniform, a pre-deceleration forward-shock emission
with a fast rise requires a homogeneous circumburst medium and a semi-relativistic
reverse-shock. These conditions may be relaxed if the ejecta kinetic energy increases
with geometrical depth, in which case the energy of the shocked gas may increase faster
than linearly with time, the forward-shock may be accelerated and the pre-deceleration
forward-shock emission may exhibit a sharp rise even for a wind-like medium or a
relativistic reverse-shock.

 The pre-deceleration forward-shock light-curves obtained for a homogeneous medium
and semi-relativistic reverse-shock are shown in the left panel of Figure \ref{s0jet}.
They exhibit the expected fast rise ($F_o \propto t^3$) until the deceleration time
$t_p \propto ({\cal E} n^{-1} \Gamma_0^{-8})^{1/3}$ when the peak flux is 
$F_p \propto {\cal E} n^{(\beta+1)/2} \Gamma_0^{4\beta}$, where ${\cal E}$ is
the ejecta kinetic energy per solid angle and $n$ the ambient medium proton density.
These dependencies show that variations of the shock energy among afterglows
induce a correlation of peak flux and epoch ($F_p \propto t_p^3$), while variations
of the ejecta initial Lorentz factor and ambient density yield anticorrelations:
$F_p \propto t_p^{-1.5\beta}$ and $F_p \propto t_p^{-1.5(\beta+1)}$, respectively.
These relations become consistent with that observed for fast-rising afterglows
(equation \ref{peak}) for $\beta \simg 1.5$ and $\beta \simg 0.5$, respectively.
However, this model cannot explain naturally the peak flux correlation with the
light-curve post-peak decay index, which would require an ad-hoc correlation of the
optical spectrum slope with the ejecta Lorentz factor or with the ambient density.

\subsubsection{Structured outflows -- off-axis observer location}

 A fast-rising afterglow light-curve can also be obtained from the forward-shock
emission if its energy is concentrated in a core seen by the observer from a location
outside the core's opening (e.g. Panaitescu, M\'esz\'aros \& Rees 1998, Granot et al
2002). In this model, the light-curve rise is caused by that, as the forward-shock is 
decelerated, its emission is less beamed relativistically and the cone of that emission 
widens.

\begin{figure}
\psfig{figure=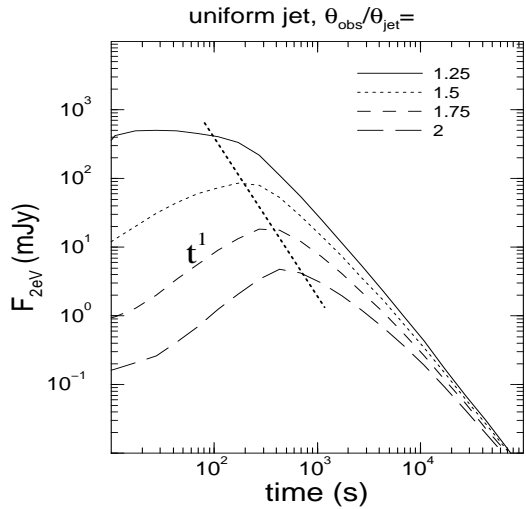,height=7cm,width=7cm} 
\caption{ 
  {\bf Forward-shock} optical light-curves for a uniform jet seen at various
   angles and a {\bf wind}-like circumburst medium. Parameters are same as for Figure
   \ref{s0jet} and the wind density is that corresponding to a stellar mass-loss rate
   to speed ratio of $10^{-7} M_\odot {\rm yr}^{-1}/({\rm km\, s^{-1}})$, which is about
   10 times denser than typical for Galactic Wolf-Rayets (a high wind density is
   required to account for the bright optical peak fluxes measured for the optical
   afterglows with a fast rise). The light-curve rise is due to the beaming cone of
   the forward-shock emission widening gradually, as the outflow decelerates.
   In contrast with the homogeneous medium (middle panel of Figure \ref{s0jet}),
   the light-curve rise is substantially slower (owing to the slower deceleration
   produced by a wind-like medium) and incompatible with observations of rising
   optical afterglows.
 }
\label{s2jet}
\end{figure}

\begin{figure}
\psfig{figure=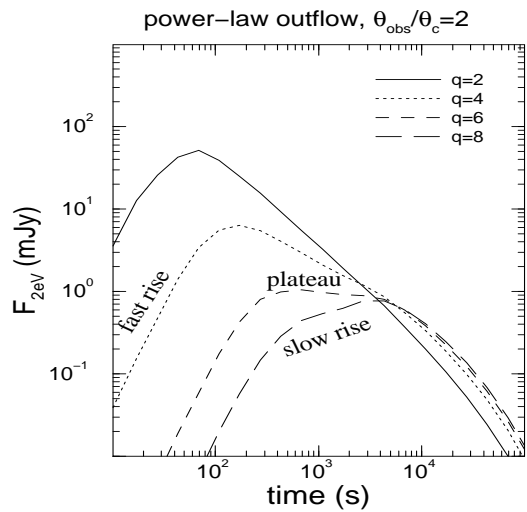,height=7cm,width=7cm} 
\caption{ 
  {\bf Forward-shock} optical light-curves for a power-law outflow interacting
  with a {\bf homogeneous} medium. The observer location is fixed ($\theta_{obs} =
  2\theta_c$) and the index $q$ of the outflow angular structure, ${\cal E} (\theta) =
  10^{54} (\theta/\theta_c)^{-q} \,{\rm erg\, sr^{-1}}$, is varied, to show its effect.
  Parameters are the same as for the right panel of Figure \ref{s0jet},
  except $\theta_c = 2$ deg. For increasing parameter $q$, the forward-shock region
  moving toward the observer has a lower kinetic energy and initial Lorentz factor,
  leading to a dimmer light-curve rise and a later peak. The shape of the optical
  light-curve also changes with $q$: early peak for $q \siml 4$, plateau for $q=6$,
  slow late rise for $q=8$.
 }
\label{qpar}
\end{figure}

 In the extreme case of a collimated outflow with a sharp angular boundary (a jet),
the light-curve rise is $F_o \propto t^4$ for a homogeneous medium (middle panel of
\ref{s0jet}) and $F_o \propto t^1$ for a wind medium (Figure \ref{s2jet}). The latter
is too slow compared to observations. A deficiency of this model is that, for observer
locations that accommodate the observed range of peak times ($t_p$) and peak fluxes
($F_p$), the resulting range of post-peak decay indices is too small.

 A wider range of post-peak decays is obtained if the outflow kinetic energy has a
smoother angular distribution then a top-hat. If the ejecta outside the core carry
a sufficiently large kinetic energy, then their pre-deceleration emission may
overshine that from the core, leading to an afterglow rise $F_o \propto t^{2-3}$
for a homogeneous external medium an a too slow $F_o \propto t^{1/2}$ for a wind.
By keeping the outflow parameters unchanged and varying the observer's location
($\theta_{obs}$, we obtain numerically that, for an outflow with a power-law angular
distribution of kinetic energy per solid angle (${\cal E} (\theta) = {\cal E}_0
(\theta/\theta_c)^{-q}$, $q >0$), the range of post-peak decay indices $\alpha_o$
increases with the structural parameter $q$. For $q=4$ and $\theta_{obs} \in (0,2)
\theta_c$, the resulting ranges of $F_p$, $t_p$ and $\alpha_o$ are compatible with
the observations of fast-rising optical afterglows (right panel of Figure \ref{s0jet}).
In the same model, a wider range of observer offsets, $\theta_{obs} \in (0,4) \theta_c$,
or a range of structural indices, $q \in (0,8)$, can account for the diversity of
early optical light-curve behaviours: fast early rises, plateaus, late slow-rises.
The dependence of the afterglow light-curve shape on the index $q$ is illustrated
in Figure \ref{qpar}).

\subsection{Reverse-shock emission}

 The reverse-shock can produce a fast-rising optical light-curve in the same ways
as the forward-shock: either through the increasing number of radiating electrons,
before the shock crosses the ejecta shell, or through the emergence of the relativistically
beamed emission from a structured outflow seen off-axis. For brevity, we consider only
the former model: the pre-deceleration emission from an isotropic outflow.

\begin{figure}
\psfig{figure=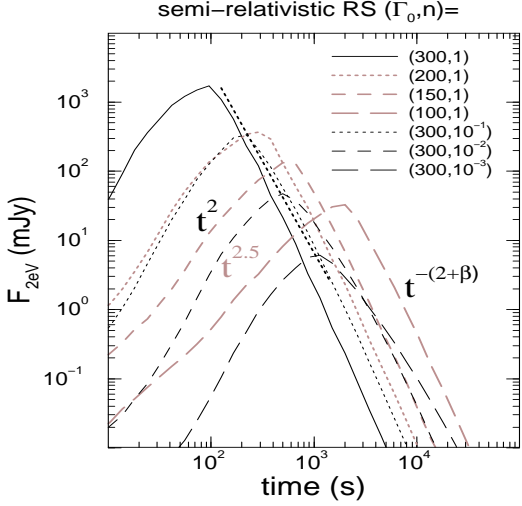,height=7cm,width=7cm} 
\caption{ 
  {\bf Reverse-shock} optical light-curves for an isotropic outflow at $z=2$ and 
  uniform ejecta (which leads to a $F_o \propto t^2$ light-curve rise).
  The ejecta are dense and the reverse-shock semi-relativistic, thus the time when it
  crosses the ejecta shell (and the light-curve peaks) is set by the ejecta Lorentz
  factor and ambient medium density (given in legend).
  The ejecta kinetic energy per solid angle is ${\cal E}_{ej}= 10^{53}\,{\rm erg\, sr^{-1}}$.
  A homogeneous medium was assumed; similar light-curves are obtained for a wind-like
  medium by changing the distribution of the ejecta kinetic energy with Lorentz factor
  or with geometrical depth in the incoming outflow.
  Microphysical parameters close to equipartition ($\varepsilon_e = 0.3$, $\varepsilon_B
  = 0.1$) were used, to obtain a sufficiently high electron synchrotron characteristic
  frequency and to match the observed optical fluxes.
  The peak flux -- peak time correlation is as observed for fast-rising optical afterglows
  if the ejecta Lorentz factor is the same but the ambient medium density varies among
  afterglows (a weaker dependence is obtained by varying the ejecta Lorentz factor).
 }
\label{rspeak}
\end{figure}

 Figure \ref{rspeak} shows the fast-rising optical light-curves obtained for a uniform
ejecta (i.e. zero radial gradient of its mass and Lorentz factor) and a semi-relativistic
reverse-shock. As for the forward-shock pre-deceleration emission, the peak flux --
peak time relation observed for the fast-rising optical light-curves is better
accommodated if the circumburst medium density is not universal.

\begin{figure}
\psfig{figure=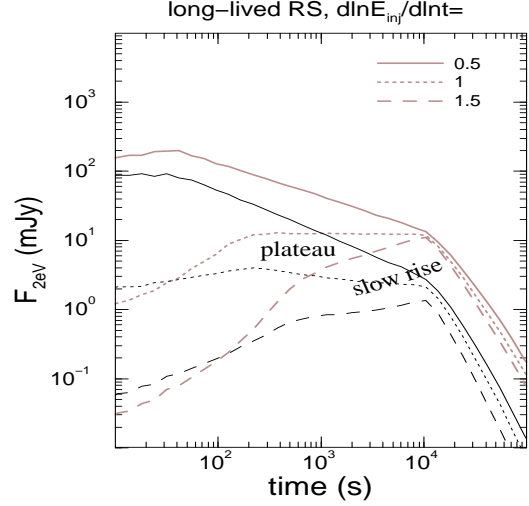,height=7cm,width=7cm} 
\caption{ 
  {\bf Reverse-shock} optical light-curves for an isotropic outflow and non-uniform 
  ejecta, whose interaction with the ambient medium increases the energy of the shocked
  fluid as a power-law of exponent given in legend, until 10 ks, when it is assumed
  that the reverse-shock crosses the ejecta shell.
  Upper set of light-curves (thick lines) are for a relativistic reverse-shock and ejecta
  initial Lorentz factor $\Gamma_0 = 1000$. Lower set of curves (thin lines) are for a
  semi-relativistic reverse-shock of variable Lorentz factor; the Lorentz factor of the
  ejecta entering the reverse-shock is determined from the kinematics of their catching-up
  with the shocked gas, assuming that all the ejecta were released instantaneously but
  with a range of Lorentz factors. The result of this set-up is that the Lorentz factor
  of the incoming ejecta is a factor $\siml 2$ larger than that of the shocked fluid.
  Other parameters are same as for Figure \ref{rspeak}.
  Note that by altering the energy injection law, reverse-shock light-curves displaying
  plateaus and slow-rises can be obtained.
 }
\label{rsflat}
\end{figure}

 The temporal behaviour of the pre-deceleration reverse-shock light-curve depends
on the radial distribution of ejecta mass and Lorentz factor. The former sets the
number of electrons accelerated by the reverse-shock and, together with the Lorentz
factor of the incoming ejecta, determines the comoving frame density of the incoming
ejecta and the Lorentz factor the reverse-shock. Figure \ref{rsflat} shows that,
for same microphysical parameters as for Figure \ref{rspeak}, reverse-shock
light-curves with plateaus and slow-rises can be obtained for non-uniform ejecta,
where that non-uniformity is quantified by the energy which they dissipate in the
shocked gas.

 Owing to that the reverse-shock is, most likely, much less relativistic than the
forward-shock during the early after phase, near equipartition magnetic fields are
required for the reverse-shock optical flux to be as high as 1 Jy. Consequently,
the cooling frequency of the reverse-shock emission spectrum is quite likely below
optical. In this case, if the injection of ejecta were to cease at some time and
the reverse shock to disappear, then the rapid electron radiative cooling would
quickly bring their synchrotron characteristic frequency below the optical and would
switch-off fast the optical emission after that time. For an isotropic outflow
(or a sufficiently wide jet), the optical emission received after the reverse-shock
disappears will be "large-angle emission" released by the reverse-shock regions moving
at angles $\theta > 1/\Gamma_0$, which arrives at observer later than the $\theta <
1/\Gamma_0$ emission and is less enhanced by relativistic beaming.

  The large-angle emission decay is $F_o \propto t^{-2-\beta}$ thus, if the peak
of the fast-rising optical light-curves were identified with the cessation of ejecta
injection in the reverse-shock, then an optical spectrum $F_\nu \propto \nu^{1/3}$
or harder would be required to accommodate the post-peak decay of the fast-rising optical
afterglows. Such spectra are much harder than the early optical spectrum of GRB afterglows
060418 and 060607A. Therefore, if the reverse-shock of an isotropic outflow is the origin
of the early optical afterglow emission, then that shock should exist and accelerate
fresh electrons even after the light-curve peak epoch $t_p$ and the peak should be
identified with a change in the radial distribution of ejecta mass and/or Lorentz at
a geometrical depth of order $ct_p$. In this model, the peak flux correlation
with the post-peak decay index found for the fast-rising optical afterglows requires
a correlation between $ct_p$ and the radial distribution of ejecta mass or Lorentz
factor at depths larger than $ct_p$. The alternative to this ad-hoc assumption is that
the fast-rising optical afterglows are not the reverse-shock pre-deceleration emission
but arise from a structured outflow seen off-axis.

\section{Conclusions}

 The early afterglow optical light-curves exhibit diverse behaviours: a third of them
display a fast or slow rise to a peak at about 100 s, a fifth have a plateau (or very
slow decay) until about 10 ks, the rest exhibiting a fast decay since first measurement.
The optical luminosity of the last two types of afterglows has a width of 2--3 dex,
while that of the fast-rising afterglows is only 0.3 dex at 0.5--5 ks after trigger,
marking the upper limit of afterglow optical luminosity (Figure \ref{z2}). 

 Additionally, the afterglows with fast rises display a good anticorrelation of the 
optical light-curve peak flux with peak epoch. This correlation extends to the afterglows 
with slow rises, suggesting that rising afterglows represent a single class. If the
dispersion in this correlation can be significantly reduced with a larger sample of 
rising afterglows, the peak flux $F_p$ - peak time $t_p$ correlation manifested by 
this class of afterglows could make it a useful standard candle. 

 With the aid of numerical calculations, we have attempted to identify models that account 
for the peculiar properties of the {\sl fast-rising} afterglows: power-law index of the 
rise, $F_p-t_p$ anticorrelation, $F_p$--decay index $\alpha_o$ correlation.

 The $F_o \propto t^{2-3}$ rise observed for those afterglows can be accommodated by either 
the pre-deceleration synchrotron emission from a relativistic blast-wave or by the emergence 
of the relativistically beamed emission from a tightly collimated outflow seen from an 
off-aperture location.

 Before deceleration, a fast brightening of the blast-wave synchrotron emission results 
from the continuous increase of the number of radiating electrons -- either the ambient 
medium electrons energized by the forward-shock or the ejecta electrons accelerated by 
the reverse-shock. For either shock, the $F_p-t_p$ anticorrelation observed for the 
fast-rising optical afterglows (Figure \ref{Fptp}, equation \ref{peak}) is more likely
to arise from variations in the circumburst medium density among afterglows (left panel 
of Figure \ref{s0jet} and Figure \ref{rspeak}).
 However, the pre-deceleration model for the optical rise does not offer a natural 
explanation for the $F_p-\alpha_o$ correlation, as the parameters which determine the
peak flux (primarily the ambient medium density) should not be correlated with those
which set the light-curve decay index depends (the distribution with energy of the 
forward-shock electrons on the radial distribution of ejecta mass and/or Lorentz factor)

 Instead, a $F_p-t_p$ relation compatible with observations is obtained for the synchrotron
emission from a collimated blast-wave and for observer locations just outside the jet
aperture (middle panel of Figure \ref{s0jet} and Figure \ref{s2jet}). In this model, 
the rise of the afterglow light-curve is due to the cone of relativistically beamed jet 
emission becoming ever wider, as the jet decelerates progressively. 
 We find that the range of post-peak light-curve decay indices measured for the fast-rising
afterglows (Table 1) is better accommodated by a structured outflow endowed with a bright 
core and a power-law distribution of the kinetic energy per solid angle in the envelope 
(right panel of Figure \ref{s0jet}).

 The optical light-curves with a {\sl decay} since the first measurement may also originate 
from structured outflows and would correspond to an observer location within the aperture of 
the brighter outflow core and a short deceleration timescale. In the same model, the afterglows 
with {\sl slow-rises} and {\sl plateaus} can be attributed to larger observer offsets relative 
to the outflow's symmetry or to the energy per solid angle decreasing away from that axis faster 
than for the fast-rising light-curves (Figure \ref{qpar}). 

 Therefore, the angular structure of the relativistic outflow and variations in the observer 
location may account for the diversity manifested by the early optical afterglow light-curves 
and the correlations displayed by the fast-rising afterglows.

\section*{Acknowledgments}
 The authors acknowledge the great help provided by the GRB data repository site
maintained by Jochen Greiner (MPE): {\sl www.mpe.mpg.de/$\sim$jcg/grbgen.html}

\end{document}